\documentclass[12pt,a4paper]{article}
\usepackage{amsmath,amssymb}
\usepackage{wasysym}
\usepackage{graphicx}
\usepackage{subfigure}
\newcommand{\wt}{\widetilde}

\newcommand{\mO}{\mathcal{O}}
\newcommand{\M}{\mathcal{M}}
\newcommand{\al}{\alpha}
\newcommand{\lb}{\left(}
\newcommand{\rb}{\right)}

\makeatletter
\newif\if@preliminary
\@preliminaryfalse
\def\preliminary{\@preliminarytrue}

\def\preprintno#1{\def\@preprintno{#1}}
\def\address#1{\def\@address{#1}}
\def\email#1#2{\thanks{\tt #1@{}#2}}
\def\abstract#1{\def\@abstract{#1}}
\renewcommand\abstractname{ABSTRACT}
\newlength\preprintnoskip
\newlength\abstractwidth
\@titlepagetrue
\renewcommand\maketitle{\begin{titlepage}%
  \let\footnotesize\small
  \hfill\parbox{\preprintnoskip}{%
  \begin{flushright}\@preprintno\end{flushright}}\hspace*{1cm}
  \vskip 60\p@
  \begin{center}%
    {\Large\bf\boldmath \@title \par}\vskip 1cm%
    {\sc\@author \par}\vskip 3mm%
    {\@address \par}%
    \if@preliminary
      \vskip 2cm {\large\sf PRELIMINARY DRAFT \par \@date}%
    \fi
  \end{center}\par
  \@thanks
  \vfill
  \begin{center}%
    \parbox{\abstractwidth}{\centerline{\abstractname}%
    \vskip 3mm%
    \@abstract}
  \end{center}
  \end{titlepage}%
  \setcounter{footnote}{0}%
  \let\thanks\relax\let\maketitle\relax
  \gdef\@thanks{}\gdef\@author{}\gdef\@address{}%
  \gdef\@title{}\gdef\@abstract{}\gdef\@preprintno{}
}%
\def\@citex[#1]#2{\if@filesw\immediate\write\@auxout{\string\citation{#2}}\fi
  \def\@citea{}\@cite{\@for\@citeb:=#2\do
    {\@citea\def\@citea{,\penalty\@m}\@ifundefined
       {b@\@citeb}{{\bf ?}\@warning
       {Citation `\@citeb' on page \thepage \space undefined}}%
\hbox{\csname b@\@citeb\endcsname}}}{#1}}
\def\citerange{\@ifnextchar [{\@tempswatrue\@citexr}{\@tempswafalse\@citexr[]}}
\def\@citexr[#1]#2{\if@filesw\immediate\write\@auxout{\string\citation{#2}}\fi
  \def\@citea{}\@cite{\@for\@citeb:=#2\do
    {\@citea\def\@citea{--\penalty\@m}\@ifundefined
       {b@\@citeb}{{\bf ?}\@warning
       {Citation `\@citeb' on page \thepage \space undefined}}%
\hbox{\csname b@\@citeb\endcsname}}}{#1}}
\long\def\@makecaption#1#2{%
  \vskip\abovecaptionskip
  \sbox\@tempboxa{#1: \emph{#2}}%
  \ifdim \wd\@tempboxa >\hsize
    #1: \emph{#2}\par
  \else
    \hbox to\hsize{\hfil\box\@tempboxa\hfil}%
  \fi
  \vskip\belowcaptionskip}
\def\fmslash{\@ifnextchar[{\fmsl@sh}{\fmsl@sh[0mu]}}
\def\fmsl@sh[#1]#2{%
  \mathchoice
    {\@fmsl@sh\displaystyle{#1}{#2}}%
    {\@fmsl@sh\textstyle{#1}{#2}}%
    {\@fmsl@sh\scriptstyle{#1}{#2}}%
    {\@fmsl@sh\scriptscriptstyle{#1}{#2}}}
\def\@fmsl@sh#1#2#3{\m@th\ooalign{$\hfil#1\mkern#2/\hfil$\crcr$#1#3$}}
\makeatother

\newcommand{\GeV}{{\ensuremath\rm GeV}}
\newcommand{\TeV}{{\ensuremath\rm TeV}}

\newcommand{\fb}{{\ensuremath\rm fb}}
\newcommand{\ab}{{\ensuremath\rm ab}}

\newcommand{\chp}{\tilde{\chi}^+}
\newcommand{\chm}{\tilde{\chi}^-}
\newcommand{\chap}{\tilde{\chi}_1^+}
\newcommand{\cham}{\tilde{\chi}_1^-}
\newcommand{\chapm}{\tilde{\chi}_1^\pm}
\newcommand{\chbp}{\tilde{\chi}_2^+}

\newcommand{\chbpm}{\tilde{\chi}_2^\pm}

\newcommand{\etal}{\textit{et al.}}
\newcommand{\ME}{\mathcal{M}}

\newcommand{\whizard}{\texttt{WHIZARD}}
\newcommand{\oMega}{\texttt{O'Mega}}
\newcommand{\feynarts}{\texttt{FeynArts}}
\newcommand{\formcalc}{\texttt{FormCalc}}
\begin{document}
 \preprintno{DESY 06-103\\SI-HEP-2006-07\\hep-ph/0607127\\[0.5\baselineskip] 
   July 2006}

\title{%
 NLO Event Generation\\for Chargino Production at the ILC
}
\author{%
 W.~Kilian\email{wolfgang.kilian}{desy.de}$^{a,b}$,
 J.~Reuter\email{juergen.reuter}{desy.de}$^b$, 
 T.~Robens\email{tania.robens}{desy.de}$^b$
}
\address{\it%
$^a$Fachbereich Physik, University of Siegen, D--57068 Siegen, Germany\\
$^b$Deutsches Elektronen-Synchrotron DESY, D--22603 Hamburg, Germany\\
}

\abstract{%
  We present a Monte-Carlo event generator for simulating chargino
  pair-production at the International Linear Collider (ILC) at
  next-to-leading order in the electroweak couplings.  By properly
  resumming photons in the soft and collinear regions, we avoid
  negative event weights, so the program can simulate physical
  (unweighted) event samples.  Photons are explicitly generated
  throughout the range where they can be experimentally resolved.
  Inspecting the dependence on the cutoffs separating the soft and
  collinear regions, we evaluate the systematic errors due to soft and
  collinear approximations.  In the resummation approach, the residual
  uncertainty can be brought down to the per-mil level, coinciding
  with the expected statistical uncertainty at the ILC.
}
%


\maketitle


\section{Introduction}

The MSSM, the minimal supersymmetric (SUSY) extension of the Standard
Model (SM), is a promising candidate for a theory of electroweak
interactions~\cite{SUSY}.  In this model, the Higgs sector is
stabilized against power divergences in radiative corrections, proton
stability suggests a discrete symmetry that provides us with a
dark-matter particle, and the renormalization-group evolution of
couplings is precisely consistent with gauge-coupling unification
(GUT) at an energy scale of the order $10^{16}\;\GeV$.

A solid prediction of the MSSM is the existence of charginos
$\chapm,\chbpm$, the superpartners of the $W^\pm$ and the charged-Higgs
$H^\pm$ bosons.  In GUT models their masses tend to be near the lower
edge of the superpartner spectrum, since the absence of strong
interactions precludes large positive renormalizations of their
effective masses.  Thus, if any superpartners are accessible in
$e^+e^-$ collisions at a first-phase ILC with c.m.\ energy of
$500\;\GeV$, the lighter chargino $\chapm$ is likely to be
pair-produced with a sizable cross section.  In many models, including
popular supergravity-inspired scenarios such as
SPS1a/SPS1a'~\cite{Allanach:2002nj}, the second chargino $\chbpm$ will
also be accessible at the ILC, at least if the c.m.\ energy is
increased to about $1\;\TeV$.  Similar arguments hold for the
neutralinos, the superpartners of neutral gauge and Higgs bosons.  The
lightest neutralino is possibly the lightest superpartner (LSP) and
therefore the dark-matter particle present in the MSSM.

The precise measurement of the chargino parameters (masses, mixing of
$\chapm$ with $\chbpm$, and couplings) is a key for uncovering any of
the fundamental properties of the MSSM that we have mentioned above.
These values give a handle for proving supersymmetry in the Higgs and
gauge-boson sector and thus the cancellation of power divergences.
Charginos decay either directly or via short cascades into the LSP,
and a precise knowledge of masses and mixing parameters in the
chargino/neutralino sector is the most important ingredient for
predicting the dark-matter content of the universe.  Finally, the
high-scale evolution of their mass parameters should point to a
particular supersymmetry-breaking scenario, if the context of a GUT
model is assumed~(cf.~\cite{SPA}).  In all these cases, a knowledge of
parameters with at least percent-level accuracy is necessary.


At the LHC, this is a difficult task since charginos are accessible
mainly in complicated decay cascades of colored superpartners with
substantial background, while direct pair-production is
suppressed~\cite{LHC}. The ILC provides much cleaner production
channels and decay signatures with low background, so the required
precision will be available at the
ILC~\cite{TESLA,Weiglein:2004hn}.  To match this experimental
accuracy, there is obvious need for theoretical predictions with
next-to-leading order (NLO) accuracy in the electroweak couplings.
The predictions have to be implemented in the simulation tools that
are actually used in the experimental analyses~(e.g. see
\cite{Allanach:2006fy}).

\begin{figure}[hbt]
\begin{center}
  \includegraphics[width=100mm]{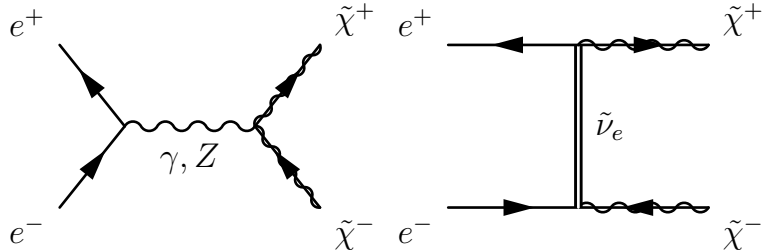}
\end{center}
\caption{Feynman graphs for chargino pair production at the ILC.}
\label{fig:LO-graphs}
\end{figure}

At leading order (LO), chargino production at the ILC is given by the
tree-level diagrams in Fig.~\ref{fig:LO-graphs}, and events can be
generated using the narrow-width approximation where all processes are
factorized in on-shell $2\to 2$ production and a cascade of on-shell
$1\to n$ decay processes.  The helicity amplitudes can be expressed in
analytic form~(cf.~\cite{ChoiChar}), and the process is available in
various computer codes~\cite{SUSY-LO}.

The NLO corrections include\footnote{This describes the multiple-pole
approximation~\cite{Denner:2000bj}; recent complete NLO calculations
in the SM~\cite{Denner:2005es} have explicitly verified the validity
of this approximation in the signal region.} (i) loop corrections to
the SUSY production and decay processes, (ii) nonfactorizable, but
maximally resonant photon exchange between production and decay, (iii)
real radiation of photons, (iv) off-shell kinematics for the signal
process, (v) irreducible background from all other multi-particle SUSY
processes, and (vi) reducible, but experimentally indistinguishable
background from Standard-Model (SM) processes.  So far, no calculation
and simulation code provides all NLO pieces for a process involving
SUSY particles.

In Ref.~\cite{CATPISS}, three computer codes have been presented and
verified against each other that simulate off-shell multi-particle
processes at tree-level, both for the SM and the MSSM.  As generators
of unweighted SUSY event samples, they thus cover (iv), (v), and (vi).
In particular, the program described in this paper is implemented as
an extension to the \whizard\ event generator~\cite{whizard}.  With
beamstrahlung, resummed initial-state radiation, arbitrary
polarization modes and standard parton-shower and had\-ron\-iz\-ation
interfaces being included, this generator is well suited for ILC
physics studies.

In this paper, we describe the extension of the tree-level simulation
of chargino production at the ILC by radiative corrections to the
on-shell process, i.e., we consider (i) in the above list and
consistently include real photon radiation (iii). This is actually a
useful approximation since in many MSSM scenarios charginos, in
particular $\chapm$, are quite narrow (cf.\
Table~\ref{tab:charginos}), so nonfactorizable contributions are
significantly suppressed and decay corrections can be separated from
the corrections to the production process.

We emphasize that for the simulation of physical (i.e., unweighted)
event samples, it is essential that the effective matrix elements are
positive semidefinite over the whole accessible phase space.  The QED
part of radiative corrections does not meet this requirement in some
phase space regions.  Methods for dealing with this problem have been
developed in the LEP1 era~\cite{LEP1}.  While these methods are also
applicable for the ILC situation, they need a thorough reconsideration
since the ILC precision actually exceeds the one achieved in LEP
experiments.

\begin{table}
  \begin{equation*}
    \begin{array}{c|cc}
      \hline
      & \text{Mass} & \text{Width} \\
      \hline
      \chap & 183.7\;\GeV & 0.077\;\GeV \\
      \chbp & 415.4\;\GeV & 3.1\;\GeV \\
      \hline
    \end{array}
  \end{equation*}
  \caption{Chargino masses and widths for the SUSY parameter set SPS1a'.}
  \label{tab:charginos}
\end{table}

\section{Fixed-Order Simulation of Chargino Production}

\subsection{Lowest Order}

In the MSSM, the charginos $\chapm,\chbpm$ are mixtures of weak
gauginos $\tilde w^\pm$ and higgsinos $\tilde h^\pm$.  The production
processes in $e^+e^-$ collisions are thus connected by SUSY
transformations to $e^+e^-\to W^+W^-$ and $e^+e^-\to H^+H^-$; the
contributing Feynman diagrams consist of $s$-channel $e^+e^-$
annihilation via $Z$ and photon and $t$-channel exchange of an
electron-sneutrino (Fig.~\ref{fig:LO-graphs}).  Since no massless
particles are exchanged in the $t$-channel, the electron mass can be
neglected at tree level throughout the phase space.  The square of the
absolute value of the matrix element, integrated over the phase space
$\Gamma$ which is parameterized by production angles $\theta,\phi$,
defines the Born cross section~$\sigma_\text{Born}$:
\begin{equation}
  \sigma_\text{Born}(s) = \int d\Gamma_2\,
  |\ME_\text{Born}(s,\cos\theta)|^2.
\end{equation}
We suppress the dependence on particle masses
$M_{\tilde\chi},M_Z,$ etc.

\begin{figure}
\begin{center}
  \includegraphics[width=120mm]{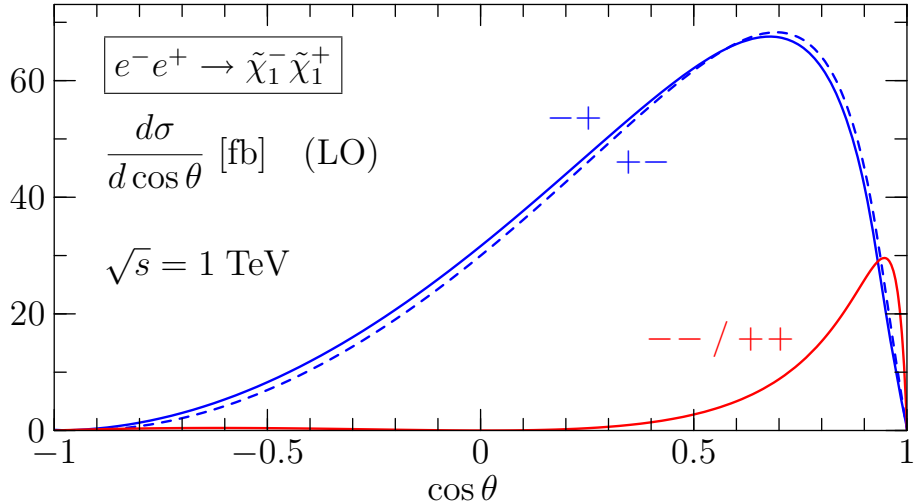}
\end{center}
\caption{Chargino pair production at the ILC: Dependence of the
  differential distribution in polar angle~$\cos\theta$ between
  $e^{-}$ and $\cham$ for different helicity
  combinations. The labels indicate $\cham$ and $\chap$ helicity;
  the electron/positron helicity is fixed to $-+$.}
\label{fig:LO-dist}
\end{figure}

At the ILC, the possibility of polarizing electrons and positrons in
the initial state will allow for separately measuring individual
(squared) helicity amplitudes.  Selecting a standard MSSM parameter
point SPS1a' (cf.~App.~\ref{app:sps} and Table~\ref{tab:charginos})
and a collider energy of $1\;\TeV$, in 
Fig.~\ref{fig:LO-dist} we display the angular dependence of the cross
section for the dominant helicity combinations with $\tilde{\nu}$
exchange in the $t$ channel.  As the amplitudes are
$\propto\,(1\,\pm\,\cos\theta), \sin\theta$ respectively, they can
become zero for $\theta\,=\,\pm\,\pi,\,0$.

\subsection{NLO Corrections}

The one-loop corrections to the process $e^-e^+\to\chm_i\chp_j$ with
$j=1,2$ have been computed in Ref.~\cite{Fritzsche}, using the
\feynarts/\formcalc\ package~\cite{FeynArts} for the evaluation of
one-loop Feynman diagrams in the MSSM.  An independent calculation
with consistent numerical results has been presented in~\cite{Wien}.
These calculations include the complete set of virtual diagrams
contributing to the process with both SM and SUSY particles in
the loop.  The collinear singularity for photon radiation off the
incoming electron and positron is regulated by the finite electron
mass~$m_e$.  As an infrared regulator, the calculation introduces a
fictitious photon mass~$\lambda$.  The interference of these diagrams
with the Born term defines the 'virtual'
contribution 
\begin{equation}
  \sigma_\text{virt}(s,\lambda^2,m_e^2) = \\ \int d\Gamma_2\left[
  2\mathrm{Re}\left(\ME_\text{Born}(s)^*\,
                    \ME_\text{1-loop}(s,\lambda^2,m_e^2)\right)\right].
\end{equation}

The dependence on the fictitious parameter~$\lambda$ is eliminated by
neglecting contributions proportional to powers of~$\lambda$ and
adding real photon radiation with energy~$E_\gamma<\Delta E_\gamma$,
where $E_\gamma$ is defined in some reference frame, usually the c.m.\
frame.  Hence, the residual logarithmic dependence on~$\lambda$ is
cancelled in favor of a logarithmic dependence on~$\Delta E_\gamma$.
This correction can be expressed as a universal factor
$f_\text{soft}(\tfrac{\Delta E_\gamma}{\lambda})$~(\ref{f-soft}).  The
`virtual + soft' contribution~$\sigma_\text{v+s}(s,\Delta
E_\gamma,m_e^2)$ is thus given by
\begin{equation}
  \sigma_\text{v+s}(s,\Delta E_\gamma,m_e^2) = \int d\Gamma_2\left[
  f_\text{soft}(\tfrac{\Delta
  E_\gamma}{\lambda})\,|\ME_\text{Born}(s)|^2 \right.  \\ 
  +
  \left. 
  2\mathrm{Re}\left(\ME_\text{Born}(s)^*\,
                    \ME_\text{1-loop}(s,\lambda^2,m_e^2)\right)\right].
\end{equation}
In a real experiment, there is always a finite energy resolution
$\Delta E_{\gamma}^\text{exp}$ for photons, and combining soft and
virtual photons below this cutoff is justified.  For the simulation
one would choose $\Delta E_\gamma \leq \Delta E_{\gamma}^\text{exp}$.

This result is complemented by the `hard' contribution
$\sigma_\text{$2\to 3$}(s,\Delta E_\gamma,m_e^2)$, i.e., the
real-radiation process $e^-e^+\to\chm_i\chp_j\gamma$ integrated over
photon phase space down to the energy resolution $\Delta E_\gamma$:
\begin{equation}
  \sigma_\text{$2\to 3$}(s,\Delta E_\gamma,m_e^2) =
  \int_{\Delta E_\gamma} d\Gamma_3\,|\ME_{2\to 3}(s,m_e^2)|^2.
\end{equation}
The sum, which can be expressed as a total cross section or, e.g., as
a differential distribution in the chargino polar angle~$\theta$, should
not depend on the photon-energy cutoff:
\begin{equation}
  \sigma_\text{tot}(s,m_e^2) = 
  \sigma_\text{Born}(s) + \sigma_\text{v+s}(s,\Delta E_\gamma,m_e^2) + 
  \sigma_\text{$2\to 3$}(s,\Delta E_\gamma,m_e^2)
\end{equation}
However, the dependence on $\Delta E_\gamma$ cancels only
approximately since positive powers of~$\Delta E_\gamma$ are neglected
in the v+s term but not in the $2\to 3$ process.

\subsection{Collinear Photons}

While photons with large energy and large angle can be experimentally
resolved and must be explicitly generated by the Monte-Carlo
simulation program, photons collinear to the incoming electrons cannot
be detected.  (Since the outgoing charginos have substantial mass, a
collinear approximation for final-state radiation is not needed.)  As
usual, we break down the (hard) $2\to 3$ cross section into a
collinear and a non-collinear part, separated at a photon
acollinearity angle $\Delta\theta_\gamma$ relative to the incoming
electron or positron:
\begin{equation}
  \sigma_\text{$2\to 3$}(s,\Delta E_\gamma,m_e^2) =
  \sigma_\text{hard,non-coll}(s,\Delta E_\gamma,\Delta\theta_\gamma)
  \\ 
  + \sigma_\text{hard,coll}(s,\Delta E_\gamma,\Delta\theta_\gamma,m_e^2),
\end{equation}
where in the non-collinear part the electron mass can be neglected.
The last term is approximated by convoluting the Born cross section with a
structure function $f(x;\Delta\theta_\gamma,\frac{m_e^2}{s})$, with
$x=1-2E_\gamma/\sqrt{s}$ being the energy fraction of the electron after
radiation,
\begin{align}
  \sigma_\text{hard,coll}(s,\Delta E_\gamma,\Delta\theta_\gamma,m_e^2)
 & = 
  \int_{\Delta E_\gamma,\Delta\theta_\gamma}d\Gamma_3\,|\ME_{2\to 3}(s,m_e^2)|^2
  \nonumber \\
 & = \int^{x_0}_{0} dx\, f(x;\Delta\theta_\gamma,\tfrac{m_e^2}{s})\int d\Gamma_2\,
  |\ME_\text{Born}(xs,m_e^2)|^2.
\end{align}
The structure function $f(x;\Delta\theta_\gamma,\frac{m_e^2}{s})$
contains two pieces $f^+,f^-$~(\ref{f-plus},\ref{f-minus}) that
correspond to helicity conservation and helicity flip, respectively;
each one is convoluted with the corresponding matrix element.  The
cutoff $\Delta E_\gamma$ is replaced by $x_0=1-2\,\Delta
E_\gamma/\sqrt s$.  In this approximation, positive powers of
$\Delta\theta_\gamma$ are neglected.

\subsection{Simulation}

Combining the above, the cross section is given by
\begin{align}\label{eq:fixed}
  \sigma_\text{tot}(s,m_e^2) &=
  \int dx\,f_\text{eff}(x_{1},\,x_{2};\Delta E_\gamma,\Delta\theta_\gamma,\tfrac{m_e^2}{s})\,
  \int d\Gamma_2\,|\ME_\text{eff}(s,x_{1},\,x_{2};m_e^2)|^2
  \nonumber \\ &\quad
  +
  \int_{\Delta E_\gamma,\Delta\theta_\gamma} d\Gamma_3\,|\ME_{2\to 3}(s)|^2,
\end{align}
where we define
\begin{align}
  f_\text{eff}(x_{1},\,x_{2};\Delta E_\gamma,\Delta\theta_\gamma,\tfrac{m_e^2}{s})
  &= \delta(1-x_{1})\,\delta(1-x_{2})
  \nonumber\\ &\quad
  + \delta(1-x_{1})\,f(x_{2};\Delta\theta_\gamma,\tfrac{m_e^2}{s})\,\theta(x_0-x_{2}) 
  \nonumber \\ &\quad
  + f(x_{1};\Delta\theta_\gamma,\tfrac{m_e^2}{s})\,\delta(1-x_{2})\,\theta(x_0-x_{1})
\end{align}
and
\begin{align}
  |\ME_\text{eff}(s,x_{1},\,x_{2};m_e^2)|^2
 & = 
    \left[1 + f_\text{soft}(\Delta E_\gamma,\lambda^2)\,\theta(x_{1},x_{2}))\right]
  \,|\ME_\text{Born}(s)|^2
  \nonumber\\ &\quad
  + 2\mathrm{Re}\left[\ME_\text{Born}(s)\,
                      \ME_\text{1-loop}(s,\lambda^2,m_e^2)\right]
  \theta(x_{1},x_{2})
\end{align}
with $\theta(x_{1},x_{2})\equiv
\theta(x_{1}-x_{0})\,\theta(x_{2}-x_{0})$.

This structure is suitable for implementing it into an event
generator.  In \whizard, for instance, there is an interface for
arbitrary structure functions $f(x_{1},x_{2})$ that can be convoluted
with the Born squared matrix element.  We insert the above effective
radiator function $f_\text{eff}$ as a `user-defined' structure
function and replace the Born matrix element as computed by the
matrix-element generator, \oMega~\cite{omega}, by the effective matrix
element defined above.  The latter is computed by a call to the
\formcalc-generated routine.

In order to account for the $\delta$-function part contained in the
radiator function, for the Monte-Carlo sampling of $x$ values the
$x_{i}$ range is separated into two regions each, one for $x_{i}<x_0$
and the other one for $x_{i}>x_0$.  For each $x_i$, the first region
is mapped such as to maximize the efficiency of event generation.  If
the sampled point ends up in the second region, $x_i$ is set equal
to~$1$ before the matrix element is evaluated as demanded by the
$\delta$ function. The relative weight of the two regions is given by 
\begin{equation}
  w(x>x_0) : w(x<x_0) = 
  1 : \int_0^{x_0}dx\,f(x;\Delta\theta_\gamma,\tfrac{m_e^2}{s}).
\end{equation}
For a consistent first-order calculation, we have to avoid the
radiation of two (collinear) photons.  Therefore, the radiator
function $f_\text{eff}$ is zero in the region with $x_1<x_0$ and
$x_2<x_0$, and in the $2\to 3$ process, no convolution with structure
functions is applied.

Implementing this algorithm in \whizard, we construct an unweighted
event generator.  With separate runs for the $2\to 2$ and $2\to 3$
parts, the program first adapts the phase space sampling and
calculates a precise estimate of the cross section.  The built-in
routines apply event rejection based on the effective weight and thus
generate unweighted event samples.  

For the $2\to 2$ part convoluted with a structure function, \whizard\
can optionally represent the missing collinear energy by a real photon
in the event, with $p_T$ generated according to the correct
logarithmic distribution up to the cutoff angle $\Delta\theta_\gamma$.
Thus, if there is any energy available for radiation, the actual
events contain a photon in addition to the chargino pair regardless
whether the event has been generated in the $2\to 2$ or $2\to 3$ part.

On the technical side, for the actual implementation we have
carefully checked that all physical parameters and, in particular, the
definition of helicity states are correctly matched between the
conventions~\cite{hagzep} used by \oMega\ and \whizard~\cite{omwhiz},
and those used by \formcalc\, (cf.~e.g.~\cite{Dittmaier:1998nn}).

\subsection{Where this Approach Fails}

Numerically, the modified \whizard\ code reproduces the total cross
section at fixed next-to-leading order in $\alpha$ as presented in
Ref.~\cite{Fritzsche}.  In principle, this makes the NLO result
available for physics simulation.
\begin{figure}
\begin{center}
  \includegraphics[width=.95\textwidth]{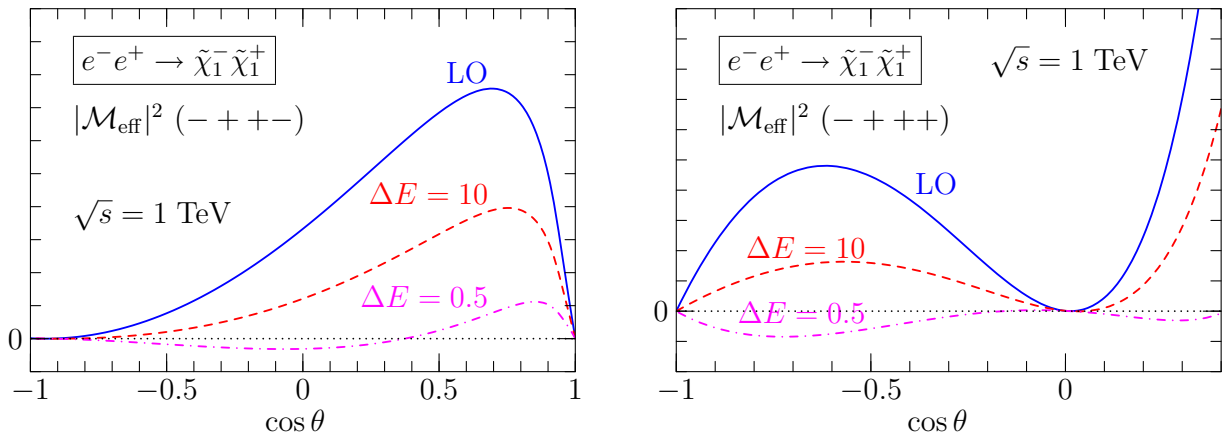}
\end{center}
\caption{Effective squared matrix element (arbitrary units) for
  $e^-e^+\to\cham\chap$ as a function of the polar scattering
  angle~$\theta$ at $\sqrt{s}=1\;\TeV$.  Left figure: Helicity
  combination $-++-$; right figure: $-+++$.  Solid line: Born term;
  dashed: including virtual and soft contributions for $\Delta
  E_\gamma=10\;\GeV$; dash-dotted: same with $\Delta E_\gamma=0.5\;\GeV$.}
\label{fig:Meff}
\end{figure}
However, in the soft-photon region the fixed-order approach runs into
the well-known problem of negative event
weights~\cite{Bohm:1986fg,k0prob}.  

While for any fixed helicity combination and chargino scattering angle
the differential cross section is positive if we include the
virtual contribution and integrate real soft photons up to a finite
cutoff~$\Delta E_\gamma$, the $2\to 2$ part of the NLO-corrected
squared matrix element is positive definite by itself only if $\Delta
E_\gamma$ is sufficiently large, say $\frac{\Delta
E_\gamma}{\sqrt{s}}=10^{-2}$ (i.e., $\Delta E_\gamma=10\;\GeV$ for
$\sqrt{s}=1\;\TeV$).  If we lower the cutoff, say $\frac{\Delta
E_\gamma}{\sqrt{s}}<10^{-3}$, for each helicity combination the
effective $2\to 2$ matrix element becomes negative within some range
of scattering angle, compensating the $2\to 3$ squared matrix element
that for such a small cutoff overshoots the total NLO differential
cross section.  This is illustrated in Fig.~\ref{fig:Meff}.

However, the experimental resolution can be better than that, and for
large $\Delta E_\gamma$ power corrections to the soft approximation
are not completely negligible; therefore we are interested in letting
$\Delta E_\gamma\to 0$.
To obtain unweighted event samples in the presence of negative-weight
events, an ad-hoc approach is to simply drop such events before
proceeding further.  The numerical consequences of such a truncation
are discussed below.\footnote{Strictly speaking, events with a weight
that can become negative do not preclude detector simulation and
further analysis, but the event samples containing them are not a
possible outcome of a physical experiment.  In particular, simulating
the detector response to such events is a waste of resources.  If a
method is available that eliminates them, it is the preferable
choice.}

Instead of slicing phase space and introducing a separate treatment of
the soft and collinear regions, an alternative approach uses
subtractions in the integrand to eliminate the singularities before
integration~\cite{Catani:1996jh}; the subtracted pieces are integrated
analytically and added back or cancelled against each other where
possible.  This method does not require technical cutoffs and
therefore allows to get rid of cutoff-induced artefacts in the result.
Unfortunately, the subtracted integrands do not necessarily satisfy
positivity conditions either, so the transformation into an unweighted
Monte-Carlo generator is not straightforward.  (Of course, the method
may be used to construct a weighted event generator.)  We do not
consider this method in the present paper.

\section{Resumming Photons}
\label{sec:resum}

\subsection{Leading Logarithms}

The shortcomings of the fixed-order approach described above are
associated with the soft-collinear region $E_\gamma<\Delta E_\gamma$,
$\theta_\gamma<\Delta\theta_\gamma$, where the appearance of double
logarithms $\frac{\alpha}{\pi}\ln\frac{E_\gamma^2}{s}\ln\theta_\gamma$
invalidates the perturbative series.  However, in that region
higher-order radiation can be resummed~\cite{softexp}.  The
exponentiated structure function
$f_\text{ISR}$ \cite{Skrzypek:1990qs}
given in Eq.~(\ref{f-ISR}) that resums initial-state radiation,
\begin{equation}
  \sigma_\text{Born+ISR}(s,\Delta\theta_\gamma,m_e^2) = 
  \\
  \int
  dx\,f_\text{ISR}(x;\Delta\theta_\gamma,\tfrac{m_e^2}{s}) \int
  d\Gamma_2\,|\ME_\text{Born}(xs)|^2,
\end{equation}
includes all-order photon radiation in the soft regime at
leading-logarithmic approximation and, simultaneously, correctly
describes collinear radiation of up to three photons in the hard
regime.  It does not account for the helicity-flip part
$f^-$~(\ref{f-minus}) of the fixed-order structure function; this may
either be added separately or just be dropped since it is subleading
in the leading-logarithmic approximation.

In this description of the collinear region, there is no explicit
cutoff $\Delta E_\gamma$ involved, and collinear virtual photons
connected to at least one incoming particle are included. The latter
part is effectively smeared over small photon energies, such that the
$+$-distribution singularity of the finite-order result is replaced by
a power-like behavior with a finite limit for $x\to 1$.  Stated
differently, the cancellation of infrared singularities between
virtual and real corrections is built-in (for collinear photons), so
that the main source of negative event weights is eliminated.

\subsection{Matching with NLO}
\label{sec:matching}

We combine the ISR-resummed LO result with the additional NLO
contributions described in the previous section.  To achieve this, we
first subtract from the effective squared matrix element, for each
incoming particle, the contribution of one soft photon that is
contained in the ISR structure function (and has already been
accounted for in the soft-photon factor),
\begin{equation}
  f_\text{soft,ISR}(\Delta E_\gamma,\Delta\theta_\gamma,m_e^2) =
  \frac{\eta}{4}\int_{x_0}^{1} dx \left(\frac{1+x^2}{1-x}\right)_+ 
  \\
  =
  \frac{\eta}{4}\left(2\ln(1-x_0) + x_0 + \frac12 x_0^2\right).
\end{equation}
Here, $\eta$ is defined in Eq.~(\ref{eq:eta}), and the $+$-distribution
is represented, e.g., by 
\begin{equation}
  g(x)_+ = 
  \\ 
  \lim_{\epsilon\to 0}
  \left[\theta(1-x-\epsilon)\,g(x) 
    - \delta(1-x-\epsilon)\int_0^{1-\epsilon} g(y)\,dy\right].
\end{equation}
After this subtraction we have
\begin{align}
  |\widetilde\ME_\text{eff}
  (\hat{s};\Delta E_\gamma,\Delta\theta_\gamma,m_e^2)|^2 & = 
  \left[1 + f_\text{soft}(\tfrac{\Delta E_\gamma}{\lambda})
           - 2f_\text{soft,ISR}(\Delta E_\gamma,\Delta\theta_\gamma,\tfrac{m_e^2}{s})\right]
  \,|\ME_\text{Born}(\hat{s})|^2
  \nonumber\\ &\quad
  + 2\mathrm{Re}\left[\ME_\text{Born}(\hat{s})\,
                      \ME_\text{1-loop}(\hat{s},\lambda^2,m_e^2)\right], 
\end{align}
with $\hat{s}$ being the c.m.\ energy after radiation.  This
expression contains the Born term, the virtual and soft collinear
contribution with the leading-logarithmic part of virtual photons and
soft collinear emission removed, and soft non-collinear radiation of
one photon; it still depends of the cutoff $\Delta E_\gamma$.  Convoluting
this with the resummed ISR structure function,
\begin{align}
  \lefteqn{\sigma_\text{v+s,ISR}(s,\Delta E_\gamma,\Delta\theta_\gamma,m_e^2)}\nonumber\\
& =
  \int dx_{1}\,f_\text{ISR}(x_{1};\Delta\theta_\gamma,\tfrac{m_e^2}{s})\,
  \int dx_{2}\,f_\text{ISR}(x_{2};\Delta\theta_\gamma,\tfrac{m_e^2}{s})
  \int d\Gamma_2\,
    |\widetilde\ME_\text{eff}(\hat s;\Delta E_\gamma,\Delta\theta_\gamma,m_e^2)|^2,
\end{align}
we obtain a modified $2\to 2$ part of the total cross section.  Note
that this convolution also includes soft and collinear photonic
corrections to the Born/one-loop interference.  The complete result
again contains in addition the $2\to 3$ part,
\begin{align}\label{eq:ISR}
   \sigma_\text{tot,ISR}(s,m_e^2) &= \sigma_\text{v+s,ISR}
  +
  \int_{\Delta E_\gamma,\Delta\theta_\gamma} d\Gamma_3\,|\ME_{2\to 3}(s)|^2.
\end{align}

\subsection{Simulation}

\begin{figure}
\begin{center}
  \includegraphics[width=.95\textwidth]{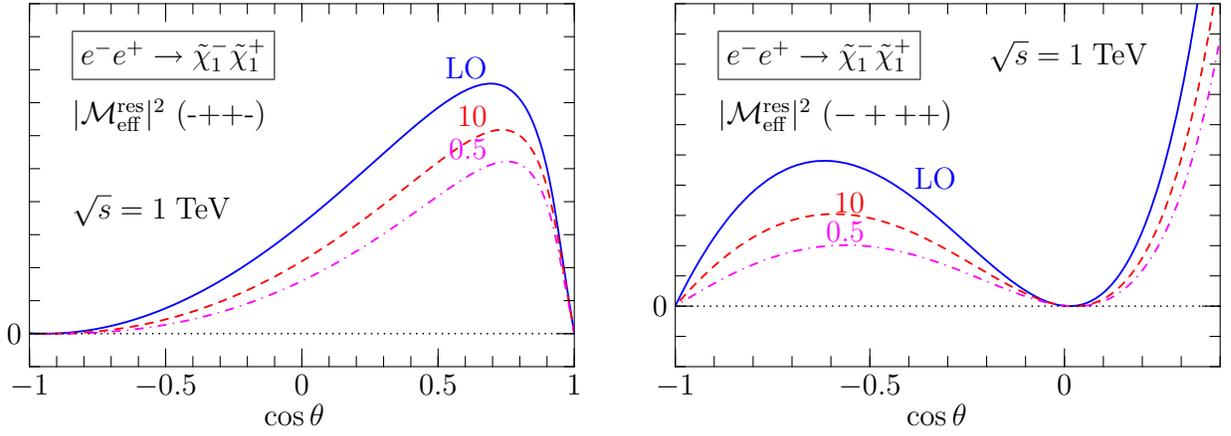}
\end{center}
\caption{Effective squared matrix element (arbitrary units) with
  the one-photon ISR part subtracted, for $e^-e^+\to\cham\chap$ as a
  function of the polar scattering angle~$\theta$ at
  $\sqrt{s}=1\;\TeV$. Left figure: Helicity combination $-++-$; right
  figure: $-+++$.  Solid line: Born term; dashed: including virtual
  and soft contributions for $\Delta E_\gamma=10\;\GeV$; dotted: same with
  $\Delta E_\gamma=0.5\;\GeV$.  The collinear cutoff is fixed at
  $\Delta\theta_\gamma=1^\circ$.}
\label{fig:Meff-ISR}
\end{figure}

As can be verified in Fig.~\ref{fig:Meff-ISR}, the resummation
approach does eliminate the problem of negative weights: shifting the
energy cutoff below the experimental resolution, e.g., $\Delta
E_\gamma=0.5\;\GeV$, such that photons are explicitly generated
whenever they can be resolved, the subtracted effective squared $2\to
2$ matrix element is still positive semidefinite in the whole phase
space.  Since neither the inclusion of the ISR structure function nor
the addition of the $2\to 3$ part introduces further sources of
negative weights, unweighting of generated events is now possible, so
this method allows for realistic simulation at NLO.

After resummation, the only potentially remaining source of negative
event weights is the soft-noncollinear region. Negative weights are
absent as long as 
\begin{equation*}
   \mathcal{O}(1) \times \frac{\alpha}{\pi} \ln \frac{(\Delta
   E_\gamma)^2}{s} \ln (\Delta \theta_\gamma) 
\end{equation*}
stays smaller than one, where the $\mathcal{O}(1)$ prefactor depends
on the specific process. For the cutoff and parameter ranges we
are considering here, this condition is fulfilled.

A final possible improvement is to also convolute the $2\to 3$ part with
the ISR structure function,
\begin{align}\label{eq:ISR+}
  \lefteqn{\sigma_\text{tot,ISR+}(s,m_e^2)}\nonumber\\
& =
  \int dx_{1}\,f_\text{ISR}(x_{1};\Delta\theta_\gamma,\tfrac{m_e^2}{s})\,
  \int dx_{2}\,f_\text{ISR}(x_{2};\Delta\theta_\gamma,\tfrac{m_e^2}{s})\,
  \nonumber\\
&\quad  \times
  \left(\int d\Gamma_2\,
    |\widetilde\ME_\text{eff}(\hat s;\Delta E_\gamma,\Delta\theta_\gamma,m_e^2)|^2
  + 
  \int_{\Delta E_\gamma,\Delta\theta_\gamma} d\Gamma_3\,|\ME_{2\to 3}(\hat s)|^2
  \right).
\end{align}
This introduces another set of higher-order corrections, namely those
where after an arbitrary number of collinear photons, one hard
non-collinear photon is emitted.  This additional resummation does not
double-count.  It catches logarithmic higher-order contributions where
ordering in transverse momentum can be applied.  Other,
logarithmically subleading contributions are missed; this is
consistent since the genuine second-order part is not calculated
anyway.

\section{Results}

\subsection{Choosing Cutoffs}
\label{sec:cutoffs}

In the kinematical ranges below the soft and collinear cutoffs,
several approximations are made.  In particular, the method neglegts
contributions proportional to positive powers of $\Delta E_\gamma$ and
$\Delta\theta_\gamma$, so the cutoffs must not be increased into the
region where these effects could become important.  On the other hand,
decreasing cutoffs too much we can enter a region where the limited
machine precision induces numerical instabilities.  Therefore, we have
to check the dependence of the total cross section as calculated by
adding all pieces and identify parameter ranges for $\Delta E_\gamma$
and $\Delta\theta_\gamma$ where the result is stable but does not
depend significantly on the cutoff values.

\begin{figure*}
\begin{center}
  \includegraphics[width=.95\textwidth]{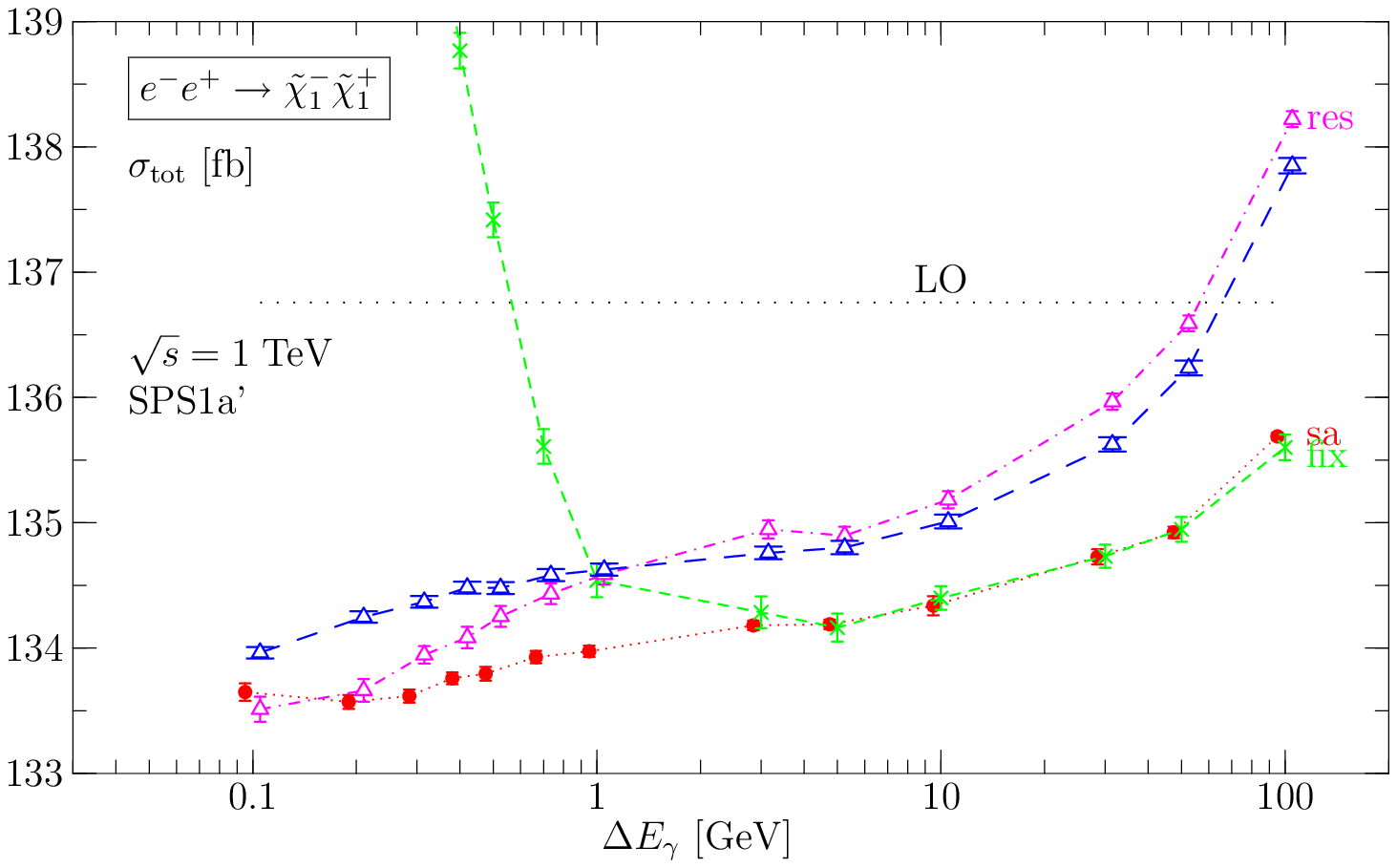}
  \vspace{\baselineskip}
  \caption{Total cross section dependence on the energy cutoff $\Delta
    E_\gamma$ using different calculational methods: {\rm `sa'}
    (dotted) = fixed-order semianalytic result~\cite{Fritzsche} using
    \feynarts/\formcalc; {\rm `fix'} (dashed) = fixed-order
    Monte-Carlo result~(\ref{eq:fixed}) using \whizard; {\rm `res'}
    (long-dashed) = ISR-resummed Monte-Carlo result~(\ref{eq:ISR+})
    using \whizard; (dash-dotted) = same but resummation applied only to
    the $2\to 2$ part~(\ref{eq:ISR}).  Statistical Monte-Carlo
    integration errors are shown.  For the Monte-Carlo results, the
    collinear cutoff has been fixed to $\Delta\theta_\gamma=1^\circ$.
    The Born cross section is indicated by the dotted horizontal
    line.}
\label{fig:edep}
\end{center}
\end{figure*}

\begin{figure*}[t]
\begin{center}
  \includegraphics[width=.95\textwidth]{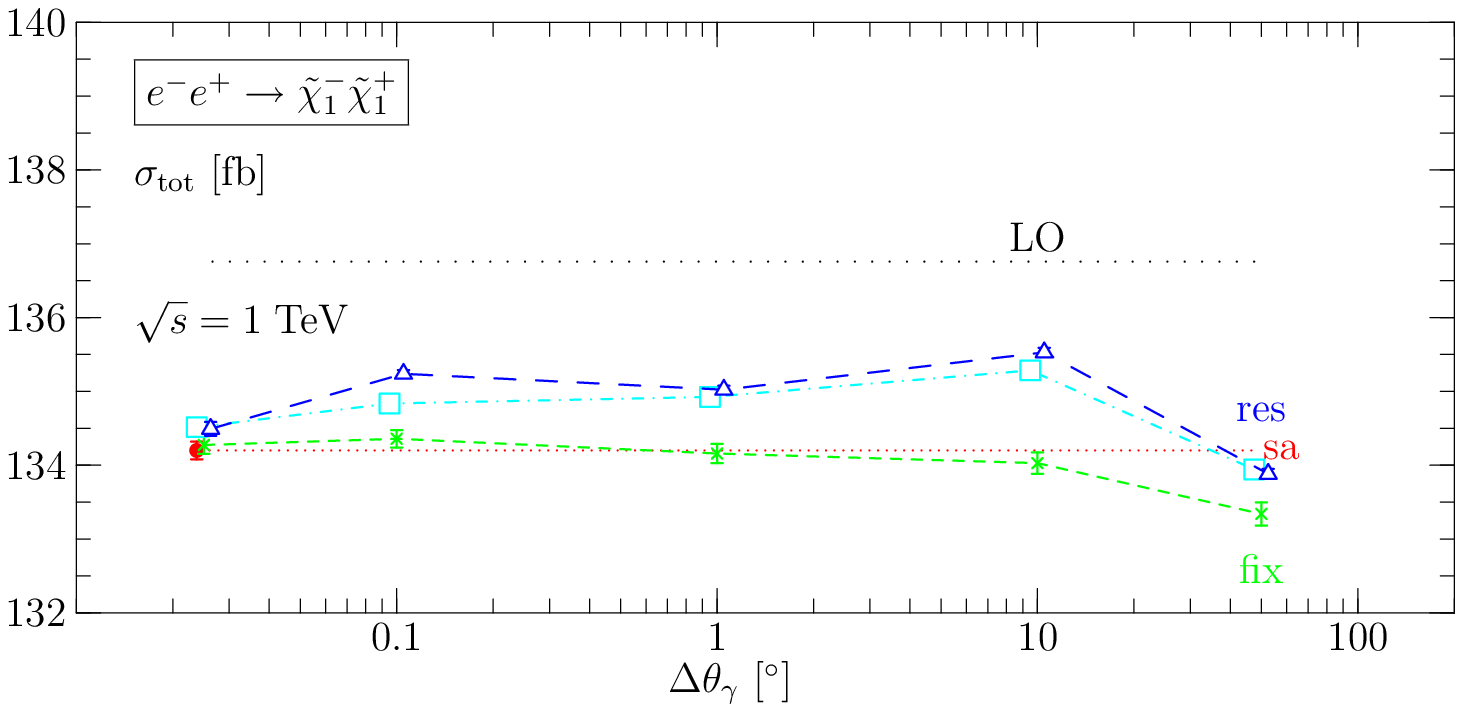}
  \vspace{\baselineskip}
  \caption{Total cross section dependence on the collinear cutoff
    $\Delta \theta_\gamma$ using three different calculational
    methods: {\rm sa} (dotted) = fixed-order semianalytic result
    \cite{Fritzsche} using \feynarts/\formcalc; {\rm fix} (dashed) =
    fixed-order Monte-Carlo result~(\ref{eq:fixed}) using \whizard;
    {\rm res} (long-dashed) = third-order ISR-resummed Monte-Carlo
    result~(\ref{eq:ISR+}) using \whizard; (dash-dotted) = same, but
    using the first-order ISR-resummed structure function instead.
    Statistical Monte-Carlo integration errors are shown.  The soft
    cutoff has been fixed to $\Delta E_\gamma=10\;\GeV$.}
\label{fig:thdep}
\end{center}
\end{figure*}

\subsubsection*{Energy cutoff dependence}

In Fig.~\ref{fig:edep} we compare the numerical results obtained using
the semianalytic fixed-order calculation with our Monte-Carlo
integration in the fixed-order and in the resummation schemes,
respectively.  Throughout this section, we set the process energy to
$\sqrt{s}=1\;\TeV$ and refer to the SUSY parameter point SPS1a'.  All
$2\to 2$ and $2\to 3$ contributions are included, so the results would
be cutoff-independent if there were no approximations involved.

The fixed-order Monte-Carlo result agrees with the semianalytic
result, as it should be the case, as long as the cutoff is greater
than a few $\GeV$.  For smaller cutoff values the Monte-Carlo result
drastically departs from the semianalytic one because the virtual
correction exceeds the LO term there, and therefore the $2\to 2$
effective squared matrix element becomes negative in part of phase
space.  For the Monte-Carlo approach, aiming at unweighting events,
the integrand is set to zero in regions where it is actually negative,
and the result overshoots when this happens.

The semianalytic fixed-order result is not exactly
cutoff-independent, as one could naively expect.  Instead, it exhibits
a slight rise of the calculated cross section with increasing cutoff;
for $\Delta E_\gamma=1\;\GeV$ ($10\;\GeV$) the shift is about
$2\,\permil$ ($5\,\permil$) of the total cross section, respectively.
While for cutoff values approaching the process energy the soft
approximation is expected to fail, the rise at small cutoff values is
due to the fact that in the soft-photon factor
$f_\text{soft}(\tfrac{\Delta E_\gamma}{\lambda})$ the kinematics is
slightly tweaked (necessary to cancel the photon-mass dependence in
the virtual part), and the cancellation of the logarithmic singularity
in $\Delta E_\gamma$ is therefore not exact.  The mismatch is
parameterically of the order $\Delta E_\gamma/\sqrt{s}$ multiplied by
the virtual correction; given the fact that the virtual correction
exceeds the LO term at about $\Delta E_\gamma=1\;\GeV$, we expect an
error of up to a few~$\permil$ at that point.  For $E\lesssim
0.1\;\GeV$, this error becomes truly negligible.

The fully resummed result~(\ref{eq:ISR+}) shows an increase of about
$5\,\permil$ of the total cross section with respect to the
fixed-order result which stays roughly constant until $\Delta
E_\gamma>10\;\GeV$ where the soft approximation breaks down.  This
increase is a real effect; it is due to higher-order photon radiation
that is absent from the fixed-order calculation.  Comparing with the
curve for $2\to 2$ resummation (\ref{eq:ISR}), we observe that for
$\Delta E_\gamma>1\;\GeV$ these higher-order contributions are caught
by ISR resummation of the $2\to 2$ part, but are transferred to the
$2\to 3$ part if the cutoff is lowered further, i.e., one radiated
photon is resolved.

At the ILC, with a cross section of more than $100\;\fb$ and an
integrated luminosity of $1\;\ab^{-1}$, a statistical $1\sigma$
fluctuation level of $2.5\,\permil$ is reached.  Although systematical
uncertainties in the analysis are likely to be relevant as well, the
theoretical prediction of the Monte-Carlo generator should aim at
matching that precision.  To get rid of artefacts of the soft
approximation at the level of $2\,\permil$, we have to choose $\Delta
E_\gamma\leq 0.5\;\GeV$.  (For cutoffs lower than $0.1\;\GeV$
double-precision numerics breaks down, although we still could obtain
results by switching to quadruple precision.)  Resumming photons leads
to an increase of $5\,\permil$. 

A common practice is to just convolute the Born part with the ISR structure
function, leaving all other contributions at fixed $O(\alpha)$.  Our
results show that this is insufficient if the achievable accuracy is
to be exploited.

To conclude, for the resummation method indicated by
Eq.~(\ref{eq:ISR+}) the desired low cutoff values are actually
acceptable, and the systematic errors induced by the soft and
collinear approximations can be suppressed down to the expected level
of statistical fluctuations. In principle, NNLO corrections and
higher order effects of running couplings should be
studied for a final verdict on the theoretical accuracy. However, we
do not expect these corrections to be significant; in particular, at
ILC energies electroweak Sudakov logarithms are still sufficiently
small~\cite{sudakov}. Off-shell and finite-width effects can be taken
into account by interfacing the results obtained here with the
multi-particle event generators presented in~\cite{CATPISS}.

\subsubsection*{Collinear cutoff dependence}

The collinear cutoff $\Delta\theta_\gamma$ separates the region where,
in the collinear approximation, higher-order radiation is resummed
from the region where only a single photon is included, but treated
exactly.  We show the dependence of the result on this cutoff is in
Fig.~\ref{fig:thdep}.

The plot shows that the main higher-order effect is associated with
photon emission angles below $0.1^\circ$.  Cutoff values between
$0.1^\circ$ and $10^\circ$ are essentially equivalent.  To achieve
this cutoff-independence, collinear terms have to be included in the
structure function beyond first order (up to third order in our case);
using the first-order ISR-resummed structure function instead would
miss some radiation at low angles $\theta_\gamma < 1^\circ$, cf.~the
small difference at $\Delta\, \theta_\gamma = 0.1^\circ$ between the first- and
third-order results in Fig.~\ref{fig:thdep}. For $\theta_\gamma>10^\circ$,
the collinear approximation breaks down.

\subsubsection*{Photon mass dependence}

The dependence on the fictitious photon mass $\lambda$ is eliminated
by implementing the soft-photon factor $f_\textrm{soft}$ before any
further manipulations are done.  Therefore, while the photon mass
remains a parameter in the matrix element code, the result does not
numerically depend on it, regardless which method has been chosen.

\subsection{Energy Dependence of the Total Cross Section}

\begin{figure}
\begin{center}
  \includegraphics[width=.95\textwidth]{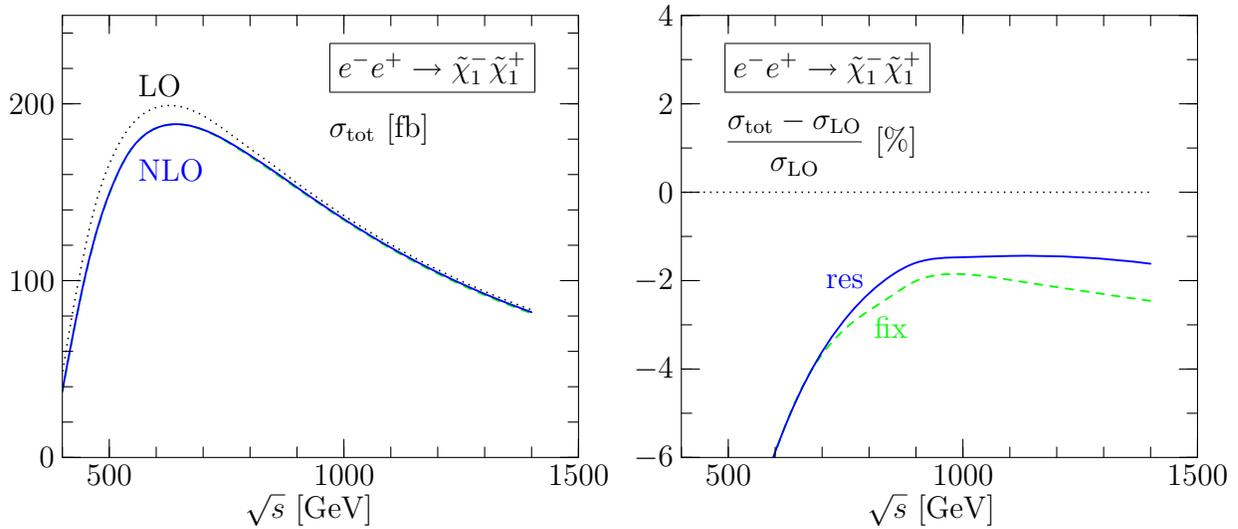}
  \vspace{\baselineskip}
  \caption{Left: $\sigma_{tot}$ in fb as a function of $\sqrt{s}$;
    right: relative correction to the Born term in percent.  LO
    (dotted) = Born cross section without ISR; NLO/fix (dashed) =
    fixed-order approach; NLO/res (full) = resummation approach.  We
    choose $\Delta E_\gamma=0.005\sqrt{s}$ and
    $\Delta\theta_\gamma=1^\circ$ as cutoffs separating the hard
    (non-collinear) from the soft (collinear) regions, respectively.}
\label{fig:sdeptot}
\end{center}
\end{figure}

Fixing the cutoffs to reasonable values, we can use the integration
part of the Monte-Carlo generator to evaluate the total cross section
at NLO for various energies.  This is shown in Fig.~\ref{fig:sdeptot},
where we display the LO result together with the NLO result for the
fixed-order and resummed approach indicated above, respectively.  Near
the cross-section maximum, the relative correction in the fixed-order
(resummed) approach is about $-5\,\%$, approaching $-2\,\%$
($-1.5\,\%$) at $\sqrt{s}=1\;\TeV$, respectively.  Near threshold and
at asymptotic energies, the relative NLO correction is larger in
magnitude.

\subsection{Simulated Distributions}

The strength of the Monte-Carlo method lies not in the ability to
calculate total cross sections, but to simulate physical event
samples.  We have used the \whizard\ event generator augmented by the
effective matrix elements and structure functions as introduced above,
to generate unweighted event samples for chargino production.

\begin{figure}
\begin{center}
  \includegraphics[width=.95\textwidth]{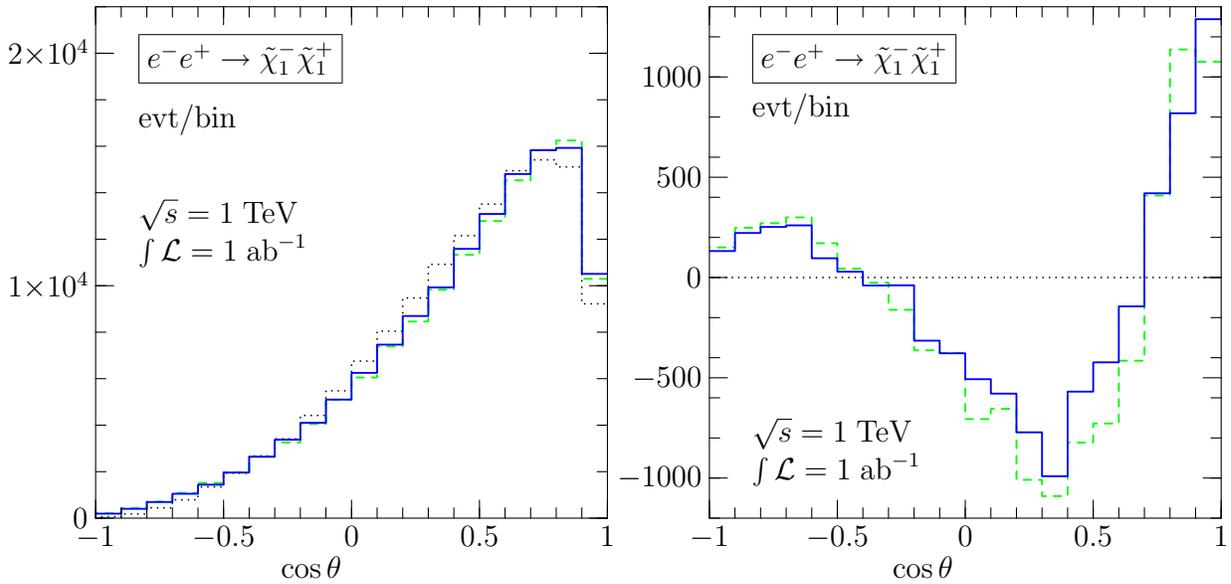}
  \vspace{\baselineskip}
  \caption{Polar scattering angle distribution for an integrated
    luminosity of $1\;\ab^{-1}$ at $\sqrt{s}=1\;\TeV$. Left: total
    number of events per bin; right: difference w.r.t.\ the Born
    distribution.  LO (dotted) = Born cross section without
    ISR; fix (dashed) = fixed-order approach; res (full)
    = resummation approach.  Cutoffs: $\Delta E_\gamma=3\;\GeV$ and
    $\Delta\theta_\gamma=1^\circ$.}
\label{fig:histth}
\end{center}
\end{figure}

To evaluate the importance of the NLO improvement, in
Fig.~\ref{fig:histth} we show the binned distribution of the chargino
production angle as obtained from a sample of unweighted events
corresponding to $1\;\ab^{-1}$ of integrated luminosity.  The collider
energy has been set to $1\;\TeV$, the SUSY parameter point
SPS1a' is the same one as for the previous plots.  With cutoffs
$\Delta\theta_\gamma=1^\circ$ and $\Delta E_\gamma=3\;\GeV$ we are not
far from the expected experimental resolution, while for the
fixed-order approach negative event weights do not yet pose a problem.
(As discussed above, for the resummation approach decreasing cutoffs
further is possible and preferred, but choosing lower values would
invalidate the fixed-order approach for the comparison.)

The histograms illustrate the fact that NLO corrections in chargino
production are not just detectable, but rather important for an
accurate prediction, given the high ILC luminosity.  The correction
cannot be approximated by a constant $K$-factor but takes a different
shape than the LO distribution.  The correction is positive
in the forward and backward directions, but negative in the central
region.  The effects of photon resummation are not as striking, but
still statistically significant; they are visible mostly in the
central-to-forward region.  Apparently, to carefully choose the
resummation method and cutoffs will be critical for a truly precise
analysis of real ILC data.

\section{Summary and Outlook}

We have presented results obtained from implementing NLO corrections
into a Monte-Carlo event generator for chargino pair-production at the
ILC.  On top of the genuine SUSY/electroweak corrections, we have
considered several approaches of including photon radiation, where a
strict fixed-order approach allows for comparison and consistency
checks with published semianalytic results in the literature, while a
version with soft- and hard-collinear resummation of photon radiation
not just improves the numerical result, but actually is more
straightforward to implement and does not suffer from negative event
weights in or near the accessible part of phase space.

A careful analysis of the dependence on the technical cutoffs on
photon energy and angle that slice phase space in regions, reveals
uncertainties related to higher-order radiation and breakdown of the
soft or collinear approximations.  For the level of precision required
by ILC analyses, the cutoffs have to be chosen rather low.
Resummation of photon radiation is required not just for precision,
but also to get rid of negative event weights in the simulation.  In
the Monte-Carlo event generator \whizard\ we have thus implemented the
NLO result with higher-order resummation in critical regions.  The
generator accounts for all yet known higher-order effects, allows for
cutoffs small enough that soft- and collinear-approximation artefacts
are negligible, and explicitly generates photons where they can be
resolved experimentally.

The generator that we have constructed should be regarded as a step
towards complete NLO simulation of SUSY processes at the ILC.  If
charginos happen to be metastable, it already provides all necessary
ingredients.  Beam effects (beamstrahlung and energy spread,
polarization) are available for simulation and can easily be included.
However, charginos are metastable only for peculiar SUSY parameter
points; in general we have to take into account chargino decay and the
corresponding additional NLO corrections.  These we have to match with
off-shell and background effects, already available for simulation in
\whizard.  Furthermore, in the threshold region the Coulomb
singularity calls for resummation, not yet accounted for in the
program.  These lines of improvement will be pursued in future
work. 


\section*{Acknowledgments}

We are grateful to T.~Hahn, W.~Hollik and T.~Fritzsche for valuable
discussions, technical help, and for providing us code prior to
publication.  We furthermore acknowledge S.~Dittmaier for helpful
discussions, and T.~Ohl for a careful reading of the manuscript.  This
work was supported by the German Helmholtz Association, Grant
VH--NG--005.

\clearpage

\begin{appendix}

\section{Soft and collinear photon factors}

The soft-photon factor that allows us to eliminate the ficticious
photon mass~$\lambda$ in favor of a physical photon-energy
cutoff~$\Delta E_\gamma$ is given by the
integral~\cite{'tHooft:1978xw,Bohm:1986fg,Denner:1991kt} 
\begin{equation}\label{f-soft}
  f_\text{soft} =
-\frac{\al}{2\,\pi}  \sum_{i,j\,=\,e^{\pm},\wt{\chi}^{\pm} } \int_{|\bf{k}|\leq\Delta E_\gamma}
  \frac{d^{3}k}{2\omega_{k}}\,\frac{(\pm) p_{i}p_{j}\,Q_{i}\,Q_{j}}{(p_{i}k)(p_{j}k)}.
\end{equation}
$p_{i}$ ($k$) denote the electron/chargino (photon)
four-vectors, respectively, while 
\begin{equation}
  \omega_{k} = \sqrt{\bf{k}^{2}+\lambda^{2}}
\end{equation}
is the energy of a photon regularized by the photon mass $\lambda$. $Q_{i}$ are the corresponding charges. The sign is $-$ for incoming and $+$ for outgoing particles, respectively.

For conserved helicity, the collinear radiation of one photon can be
approximated by convoluting the no-photon matrix element with the
structure function (see, e.g.,~\cite{Dittmaier:1993jj})
\begin{equation}\label{f-plus}
  f^+(x) = \frac{\eta}{4}\,\frac{1+x^{2}}{(1-x)}.
\end{equation}
Here, with $k^\top_{max} = p_e^0 \, \Delta\theta_\gamma$ being the
collinear cutoff, the expansion parameter~$\eta$ is defined as
\begin{equation}
  \label{eq:eta}
  \eta = \frac{2\alpha}{\pi}\left[\ln
    \left(\frac{s}{4\,m_e^2}\,(\Delta\theta_\gamma)^{2}\right)-1\right].
\end{equation}
The helicity-flip structure function,
\begin{equation}\label{f-minus}
  f^-(x) = \frac{\alpha}{2\pi} (1-x)
\end{equation}
does not contain a logarithmic enhancement with $\Delta\theta_\gamma$,
so the helicity-flip part of collinear radiation is subdominant.

In the soft-collinear region $x\approx 1$, $\theta_\gamma\approx 0$,
the leading logarithms in $\Delta\theta_\gamma$ (i.e., powers of
$\eta$) can be resummed to all orders~\cite{softexp}.  Matching this
with the complete $x$-dependent expressions for hard collinear
radiation to first, second, and third order in~$\eta$, Skrzypek and
Jadach obtained the ISR structure function~\cite{Skrzypek:1990qs}
\begin{align}\label{f-ISR}
 f_\text{ISR}(x) &=
  \frac{\exp\left(-\tfrac12\eta\gamma_{E} + \tfrac{3}{8}\eta\right)}
       {\Gamma(1+\frac{\eta}{2})}\,
  \frac{\eta}{2}\,(1-x)^{(\frac{\eta}{2}-1)}
  - \frac{\eta}{4}\,(1+x)
\nonumber\\ 
  &\quad
  + \frac{\eta^{2}}{16}\left(
      -2(1+x)\ln(1-x)
  - \frac{2\ln x}{1-x}  + \frac32(1+x)\ln x
      - \frac{x}{2} - \frac{5}{2}\right)
\nonumber\\
  &\quad
  + \frac{\eta^3}{8}\left[
       -\frac{1+x}{2} \left(
          \frac{9}{32} - \frac{\pi^{2}}{12} + \frac{3}{4}\ln(1-x)
          + \frac{1}{2}\ln^{2}(1-x) - \frac{1}{4}\ln x\ln(1-x)
\right.\right.
\nonumber\\
  &\quad\left. \qquad\qquad\qquad\quad
          + \frac{1}{16}\ln^{2} x - \frac{1}{4}\mathrm{Li}_{2}(1-x)\right)
\nonumber\\
  &\quad\qquad\quad
       + \frac{1+x^{2}}{2(1-x)}\left(
          - \frac{3}{8}\ln x + \frac{1}{12}\ln^{2} x 
          - \frac{1}{2}\ln x \ln (1-x)\right)
\nonumber\\
  &\quad\qquad\left.\quad
       - \frac{1}{4}(1-x)\left(\ln(1-x)+\frac{1}{4}\right)
       + \frac{1}{32}(5-3x)\ln x\right].
\end{align}

\section{Two-Photon Phase-Space Regimes in the Resummation Method}

The class of processes we are considering exhibits the usual infrared
and collinear (for $m_e=0$) singularities in individual contributions
to the physical result, which we treat by standard methods.  In the
fixed-order approach where only one photon is present, the
consequences of this phase-space slicing are evident: (i) In the soft
region, the photon energy is set to zero in the matrix element, and
the real contribution is cancelled against the IR-divergent part of
the virtual correction, neglecting corrections proportional to the
photon energy.  (ii) In the hard-collinear region, the photon-emission
part of the matrix element is replaced by a structure function,
neglecting corrections proportional to the separation angle.  As long
as the problem of negative event weights can be ignored, and numerics
is not an issue, the description at fixed NLO always improves if any
of those cutoffs are lowered.

However, the resummation method described in Sec.~\ref{sec:resum}
involves all orders of soft-photon radiation.  Shifting cutoffs
changes the type of higher-order contributions that are included, so
lowering cutoffs not necessarily improves the description.

To clarify this issue, let us focus on the $O(\alpha^2)$ correction,
i.e., two photons (real or virtual).  Since we do not consider
two-loop diagrams, this correction is not completely accounted for,
but dominant contributions are included.

In the resummation method, there are three different ways of dealing
with real and virtual photons: 
\begin{itemize}
\item[(a)]
soft approximation \cite{Denner:1991kt}:\\
describes collinear and non-collinear soft photons; neglects
contributions $\propto\,\frac{\Delta E_\gamma}{\sqrt{s}}$; is combined
in the sequel with the soft photonic part of the one-loop matrix element  

\item[(b)]
ISR \cite{Skrzypek:1990qs}: \\
describes collinear real and virtual photons; neglects interference
terms in photon emissions. Assumes $k^\top$-ordering of the emitted
photons, i.e. for $j\,>\,i$: $k^\top_{j}\,>\,k^\top_{i}$, and in
$n^{th}$ order: $\sum_{i=1}^n \,k^\top_{i}\,<\,k^\top_{max}$,
where $k^\top_{max}$ is fixed.

\item[(c)]
real emission given by exact (hard non-collinear) matrix element 
$\M_{2\,\rightarrow\,3}$
\end{itemize}

Considering now the treatment of two photons
(i.e. $\mathcal{O}(\alpha^2)$ corrections), at least one of the
photons is always described by the ISR structure function. But when
the Born term is convoluted with the ISR function, there are also two-photon
contributions described solely by the ISR. We have to distinguish
between the cases where (i) the two photons are attached to the same
or (ii) to different incoming particles. In case (i), we consider the
three terms
\begin{equation}
  \label{caseone}
  \mO(\al^{2})_\text{ISR} \, - \, \mO(\al)_\text{ISR}\,\mO(\al)^\text{soft}_\text{ISR} \,
  + \, \mO(\al)_\text{ISR}\,\mO(\al)_\text{soft}.   
\end{equation}
The first term contains all pairs of collinear photons from the ISR,
$k^\top$-ordered; the last term contains a first photon from ISR and a
second one from the soft-photon factor (SPF, which in the following is
understood to include the soft photonic one-loop contribution). The
term in the middle is the subtraction to avoid double-counting of soft
photons described in subsection~\ref{sec:matching}. Here both photons
are from the ISR, the first one with arbitrary energy, the second one
soft.  

If the second of the considered photons is soft, and both are
$k^\top$-ordered, then there is an exact cancellation between the
first two terms. For non $k^\top$-ordered photons, the first term
gives no contribution, and there is a cancellation between the second
and third term, which results in a difference between the ``exact''
SPF expression and the ISR LLA term for the incoming particle~$\#j$:

\begin{equation*}
  \Delta_j = \mO(\al)_{j,\text{soft}} - \mO(\al)_{j,\text{ISR}}.
\end{equation*}
In the case (ii), we write the terms schematically as 
\begin{align}
  \label{casetwo}
  & \;\; \mO(\al)_{1,\text{ISR}} \mO(\al)_{2,\text{ISR}}  \notag \\ 
  & \; + \mO(\al)_{1,\text{ISR}}\,\lb\mO(\al)_{2,\text{soft}} - 
  \mO(\al)_{2,\text{ISR}}^\text{soft}\rb \notag \\ 
  & \;\; + \lb\mO(\al)_{1,\text{soft}} -
  \mO(\al)^\text{soft}_{1,\text{ISR}}\rb\mO(\al)_{2,\text{ISR}}.
\end{align}
Since here there are always two different structure functions
involved, $k^\top$-ordering is absent, and after a cancellation of
soft terms one is left with
\begin{equation*}
  \Delta_1  \mO(\al)_{2,\text{ISR}} + \mO(\al)_{1,\text{ISR}} \Delta_2 +
  \mO(\al)_{1,\text{ISR}} \mO(\al)_{2,\text{ISR}},
\end{equation*}
which is up to the missing terms $\Delta_1 \Delta_2$ equivalent to an
SPF description for both legs.

We now investigate the changes induced by raising one of the two
cutoffs. Generally, for the contributions with two real photons, we
loose contributions if we lower the cutoffs since double photon
radiation is not accounted for by the $2\to 3$ matrix element.  Convoluting
also the $2\to 3$ matrix element with the ISR structure function as
proposed in Equ.~(\ref{eq:ISR+}) gives contributions with a collinear
and hard non-collinear photon and cures the problem. But still,
raising the cutoff $\Delta \theta_\gamma$ opens up phase space for the
first photon. Thus, raising the cutoffs gives a better description of
these contributions as long as the collinear approximation is valid.
The same holds for the energy cutoff, but we see from
Figs.~\ref{fig:edep} and~\ref{fig:thdep} that the soft approximation
fails much earlier than the collinear description.

By including the next order of real photon radiation explicitly,
i.e., 
\begin{equation*}
    \int_{\Delta E_{\gamma,i},\Delta\theta_{\gamma,i}}
    d\Gamma_4\,|\ME_{2\to 4}(s)|^2, 
\end{equation*}
we can further improve the description of this part of two-photon
phase space.  This contribution, which is however tiny for the cutoff
values considered here, can easily be added using \whizard\ as a
tree-level event generator.

For completeness, we finally discuss the reshuffling of contributions
in the overlap region of the soft-collinear and hard-collinear (ISR)
descriptions. If we raise $\Delta E_\gamma$ while keeping $\Delta
\theta_\gamma$ fixed, photons that have been hard now become soft.  In
the case (ii) (photons radiated from two different external
particles), a photon which has been described by the structure
function, comes now with the SPF. For the case (i), we have to
distinguish whether the two photons are $k^\top$-ordered or not. If
they are, the description again changes from the ISR to the SPF. If
there is no $k^\top$-ordering, then the photons either change from
hard+soft to soft+soft, which is a smooth transition where only the
last two terms of Equ.~(\ref{caseone}) are involved, or the second
photon changes to soft for the combinations hard+hard or soft+hard. In
that case there appear new contributions of the form $ \Delta
\mO(\al)_\text{ISR}$, which have not been there before.

Raising $\Delta \theta_\gamma$, while keeping $\Delta E_\gamma$ fixed, 
shuffles photons from a non-collinear to a collinear
description. The interesting region is for photons that lie in the
soft regime near the limit of the soft-collinear regime and change
into the latter after raising the angle cutoff. For $k^\top$-ordered
photons the cross-over is smooth and the second photon stays with the
SPF, while for non-ordered photons the description switches from the
SPF to the the difference between ISR and SPF. For the case (ii) of
radiation from different legs, the description always remains with the
SPF.

For more details see~\cite{trobens}.


\section{SPS1a'}
\label{app:sps}

The SUSY parameter point SPS1a' is defined in
Ref.~\cite{SPA}; it is a SUGRA-type scenario
derived from the parameter set
\begin{gather}
  m_0 = 70\;\GeV, \quad
  m_{1/2} = 250\;\GeV, \quad
  \tan\beta = 10,\quad
\nonumber\\
  \mu>0,\quad
  A_0 = -300\;\GeV.
\end{gather}
The precise spectrum and coupling parameters are computed using the
renormalization-group evolution code of Ref.~\cite{Porod:2003um}; the
values can be found in Ref.~\cite{SPA}. The chargino masses and widths are
also listed in Table~\ref{tab:charginos}.

\end{appendix}


\end{document}